\documentclass[aps,prb,twocolumn]{revtex4-1}

\usepackage{amsmath}
\usepackage{amsthm}

\usepackage{bbm,graphicx,bm,hyperref}   
\usepackage[caption=false]{subfig}

\makeatletter
\newcommand*{\defeq}{\mathrel{\rlap{%
                     \raisebox{0.3ex}{$\m@th\cdot$}}%
                     \raisebox{-0.3ex}{$\m@th\cdot$}}%
                     =}
\makeatother
\def\ave#1{\langle #1 \rangle}
\def\ii{{\rm i}}
\def\sx{\sigma^{\rm x}}
\def\sy{\sigma^{\rm y}}
\def\sz{\sigma^{\rm z}}
\def\tr#1{{\rm tr}(#1)}
\def\1{\mathbbm{1}}
\def\cL{{\cal L}}

\def\cLd{{\cal L}_{\rm dis}}

\def\cLz{{\cal L}_0}
\def\cLe{{\cal L}_{1}}
\def\cLL{{\cal L}_{\rm L}}

\def\ex#1{{\rm e}^{#1}}
\def\z{{\rm z}}
\def\Deq{D_{\rm eq}}
\def\inn#1#2{{\langle #1,#2 \rangle}}

\def\tit#1{{\em #1},}
\def\etal#1{#1}

\begin{document}

\title{Nonequilibrium steady-state Kubo formula: equality of transport coefficients}

\author{Marko \v Znidari\v c}
\affiliation{Physics Department, Faculty of Mathematics and Physics, University of Ljubljana, 1000 Ljubljana, Slovenia}
\affiliation{Adbus Salam ICTP, Strada Costiera 11, 34151 Trieste, Italy}

\date{\today}

\begin{abstract}
We address the question of whether transport coefficients obtained from a unitary closed system setting, i.e., the standard equilibrium Green-Kubo formula, are the same as the ones obtained from a weakly driven nonequilibrium steady-state calculation. We first derive a nonequilibrium Kubo-like expression for the steady-state diffusion constant expressed as a time-integral of either a current or a conserved density nonequilibrium correlation function. This expression has certain advantages over the equilibrium Green-Kubo formula, but is not clear if it gives the same value of the diffusion constant. We then rigorously show that, if the unitary dynamics is diffusive the nonequilibrium formula indeed gives exactly the same transport coefficient. The form of finite-size correction is also predicted. Theoretical results are verified by an explicit calculation of the diffusion constant in several interacting many-body models.
\end{abstract}




\maketitle

\section{Introduction}
Transport of conserved quantities is one of the simplest manifestations of nonequilibrium physics. Depending on the dynamics transport may range from ballistic (zero bulk resistance) to diffusive (finite resistance per length), over to localization (infinite resistance), or, in principle, anything in-between these extremes, usually dubbed anomalous transport. Our experience tells us that in general transport is diffusive and described by a phenomenological Fourier's law~\cite{Fourier} (or analogous Fick's, Ohm's, etc., law for other conserved quantities), however, starting from a microscopic Hamiltonian showing that is anything but simple. In particular, in one-dimensional systems transport is often not diffusive -- there can be strong effects due to low dimensionality as well as integrability that typically causes ballistic transport. Understanding transport in one-dimensional systems of interacting particles has a long history, going back to the celebrated Fermi-Pasta-Ulam-Tsingou numerical experiment~\cite{FPUT,PT}, and even today it is still very much an open problem of high interest~\cite{Joel00,Buchanan05}.

On a theoretical level one can use the Green-Kubo linear response formula to express transport coefficients in terms of the equilibrium autocorrelation function of the respective current~\cite{Pottier}. However, calculating the time-dependent correlation function is often too involved even for in principle solvable systems (such as, e.g., a Bethe ansatz solvable XXZ spin chain). Furthermore, the Green-Kubo formula involves two limits that have to be taken in the correct order (which is in practice difficult), first the thermodynamic limit (TDL), and then the limit of infinite times. One therefore has to resort to numerical calculations. To that end two different frameworks are used: (i) closed Hamiltonian evolution calculating either the equilibrium current autocorrelation function, or spreading of inhomogeneous states, and (ii) direct simulation of a nonequilibrium steady state (NESS) transport by explicitly taking into account driving reservoirs at different potential. For classical systems there are plenty of different reservoirs available (e.g., Langevin, stochastic, Nose-Hoover, etc.) and a NESS approach is the dominant one~\cite{Lepri03,Dhar08}. In the quantum domain efficiently describing reservoirs is trickier, one way is using the Lindblad master equation~\cite{Lindblad1,Lindblad2} which is though in general difficult to solve. Therefore, traditionally a unitary closed system setting has been prevalent~\cite{Peter04,Fabian07}. With the recent development of matrix-product based methods~\cite{Schol11} things are changing as direct NESS simulations of certain Lindblad master equations are efficient and are thus becoming indispensable~\cite{Saito02,Michel04,JSTAT09,Robin09,Arenas:13,Misguich:13,Schwarz16,PRL16,Poletti:18}, especially when large 1D systems are required. A pressing question therefore is whether the Hamiltonian and NESS approaches give the same transport coefficient? We stress that even for weak nonequilibrium driving the resolution is far from obvious -- on a formal mathematical level the expressions are completely different and no rigorous connection is known~\cite{Joel00} neither for classical nor for quantum systems. Furthermore, sometimes concern is expressed that an explicit driving could modify transport properties, or, that the often used boundary driving is ``unrealistic''. Due to the increasingly widespread use of Lindblad equations in transport studies resolving this question are not just of fundamental~\cite{Joel00} but also of immediate practical importance.

We address the relation between ``equilibrium'' and NESS transport coefficient in 1D quantum systems, specializing in particle transport at high temperature, where derivations are the simplest. We obtain two main results. First, we derive a NESS Kubo-like formula for the transport coefficient in a form that is useful in itself. Second, we use this formula to make a comparison with the Green-Kubo formula, showing in full generality that, provided the unitary (Hamiltonian) dynamics is diffusive, the two approaches give the same transport type and in particular the same diffusion constant. Theoretical results, which also predict a particular convergence with system size $L$, are verified in explicit many-body interacting models. 

\section{The setting}
A common way to account for an explicit coupling to reservoirs is by an appropriate master equation. Any quantum evolution should preserve the positivity of density matrices as well as its trace. If one in addition assumes that the reservoir is infinite and fast, i.e., induces a Markovian evolution, one is led to the Lindblad master equation~\cite{Lindblad1,Lindblad2}
\begin{eqnarray}
\frac{{\rm d}\rho}{{\rm d}t}&=&\cL(\rho)=\ii[\rho,H]+\cLd(\rho),
\label{eq:Lin}
\end{eqnarray}
where $\cLd(\rho)=\sum_k 2L_k \rho L_k^\dagger-\rho L_k^\dagger L_k-L_k^\dagger L_k \rho$ is a dissipator that depends on a set of Lindblad operators $L_k$. Transport properties are determined by the scaling of the current in the NESS. For weak driving we can write the Lindbladian as a sum of two linear operators
\begin{equation}
\cL=\cLz+\mu \cLe,
\end{equation}
where $\mu$ is some small parameter and $\cLz$ is Lindbladian. The (unique) steady state of $\cLz$ is denoted by $\rho_0$, $\cLz \rho_0=0$. For small $\mu$ we look for a perturbative NESS solution $\rho=\rho_0 + \mu \rho_1+\cdots$, obtaining the well known linear correction $\cLz \rho_1=-\cLe \rho_0=:-R$. Formally, one can write $\rho_1=-\cLz^{-1}(R)$. This expression has a unique solution provided $R$ is orthogonal to the kernel of $\cLz$. Alternatively, one can do a time-dependent perturbation theory (see Appendix~\ref{app1}), arriving at~\cite{Michel04}
\begin{equation}
  \rho_1=\rho_1(t \to \infty)=\int_0^\infty \ex{\cLz \tau}R{\rm d}\tau=\int_0^\infty R(\tau){\rm d}\tau.
  \label{eq:rho1R}
\end{equation}

\section{NESS Kubo}
In transport studies one often employs Lindblad operators that act only at the chain boundaries~\cite{foot0}, arguing that in the TDL~\cite{foot1} and for a self-thermalizing system~\cite{foot2} the precise form of driving should not matter for bulk physics, i.e., far away from boundaries. A popular choice, both due to the existence of exact solutions~\cite{Prosen11} as well as frequent efficiency of numerical MPS-based methods~\cite{Schol11} enabling simulation of 1D quantum systems of several hundred sites, is to take $L_j$ that act only on the system's boundary. In order to be able to execute all the steps of our derivation explicitly without any further assumptions we shall focus on the simplest and also the most common case~\cite{Michel03,Michel04,Wichterich07,JSTAT09,Popkov12,Kamiya13,Landi:14,PRL16,PRL11,PNAS18,Arenas:13,Poletti:18} of particle (magnetization) driving where one uses Lindblad operators $L_1=\sqrt{\Gamma} \sqrt{1+\mu}\sigma^+_1$, $L_2=\sqrt{\Gamma} {\sqrt{1-\mu}} \sigma^-_1$, $L_3 = \sqrt{\Gamma} \sqrt{1-\mu}\sigma^+_L$, $L_4=\sqrt{\Gamma} \sqrt{1+\mu} \sigma^-_L$. $\Gamma$ is the coupling strength while $\mu$ is the driving strength. The dissipator at the left edge acts on boundary Pauli matrices as: $\cLL(\sx_1)=-2\Gamma\sx_1$, $\cLL(\sy_1)=-2\Gamma \sy_1$, $\cLL(\sz_1)=-4\Gamma\sz_1$, $\cLL(\1_1)=4\Gamma\mu \sz_1$, and similarly with a reversed sign of $\mu$ at the right end. The unique steady state of such a 1-site dissipator is $\sim \1+\mu \sz$, i.e., driving tries to impose magnetization $+\mu$. Together with $H$ that conserves total magnetization such a Lindblad equation can be used to study high-temperature magnetization transport in many-body systems -- a question of high interest, see e.g.~\cite{Fabian07,sirker,bruno,doyon,Ours17,bulchan,Poletti:18} (using Jordan-Wigner transformation it is equivalent to particle transport).

For weak driving we split $\cL$ into an equilibrium Lindbladian $\cLz:=\cL(\mu=0)$ (the steady-state of $\cLz$ is an infinite temperature state $\rho_0 \sim \mathbbm{1}$) and perturbation $\mu \cLe:=\cL-\cLz$ (such decomposition is exact, there are no higher order terms in $\mu$). To get $\rho_1$ we need $R=\cLe(\rho_0)=4\Gamma(\sz_1-\sz_L)$. Here we explicitly see that $R$ is indeed orthogonal to the kernel of $\cLz$. For small $\mu$ the NESS expectation value of any traceless $A$ is (\ref{eq:rho1R}),
\begin{equation}
  \ave{A}=4\Gamma\mu \int_0^\infty \tr{A \ex{\cLz t} (\sz_1-\sz_L)}{\rm d}t.
  \end{equation}
We remark that the limit of small $\mu$ is (always) well behaved in a sense that the convergence radius is finite (typically large) in the TDL.

In cases when $H$ is reflection symmetric, $PHP^\dagger=H$, with $P$ being a reflection of site $k$ around the midpoint, $k \to L+1-k$, the full $\cLz$ is as well, and so we can further desymmetrize and write $\rho_1=\tilde{\rho}_1-P\tilde{\rho}_1 P^\dagger$, where $\tilde{\rho}_1:=-4\Gamma\cLz^{-1}(\sz_1)=4\Gamma \int_0^\infty \sz_1(t){\rm d}t$ and $\sz_1(t):=\ex{\cLz t}\sz_1$.  In particular, the NESS current is odd under $P$ and so the contributions from the $\sz_1$ and $\sz_L$ are the same, and one has $j=8\Gamma\mu \int_0^\infty \tr{j_{k,k+1} \ex{\cLz t}\sz_1}{\rm d}t$ (due to the continuity equation it is independent of $k$). The diffusion constant $D$ is defined via a Fick's law relation in the NESS,
\begin{equation}
  j=-D \frac{\z_L-\z_1}{L},\qquad D:=L \frac{j}{\z_1-\z_L},
\label{eq:NESSF}
\end{equation}
where $\z_k:=\tr{\rho \sz_k}$ is the NESS expectation of magnetization. Besides the current we therefore also need the boundary magnetization. Provided the system is not ballistic, such that the NESS current decays to zero in the TDL, one will have $\z_1 \to \mu$ and $\z_L \to -\mu$. To see that one writes the NESS condition at the boundary: taking $\rho\sim \1 +(\sum_k \z_k \sz_k+\frac{j}{8} \sum_k j_{k,k+1}+ \cdots)$, we get for our magnetization driving the exact stationary condition $\cL(\rho)=0=[4\Gamma\mu-4\Gamma \z_1-j]\sz_1+\cdots$, where the dots represent terms orthogonal to $\sz_1$; the three terms in the bracket that in the NESS must sum to zero come from the injection of magnetization ($\cLL(\1)$), absorption ($\cLL(\sz_1)$), and continuity equation (current flowing from the 1st site due to $[j_{1,2},H]$), respectively. We have an exact relation (independent of the details of $H$ and the value of $\mu$) $4\Gamma(\mu-\z_1)=j$, and $4\Gamma(\mu+\z_L)=j$. These relations show that, provided $j \to 0$, one has $\z_1 \to \mu$ and $\z_L \to -\mu$. Therefore, in the TDL $\z_1-\z_L \to 2\mu$ and one can write a Kubo-like NESS expression (see Ref.~\cite{Dhar11} for classical heat conduction and Ref.~\cite{Kamiya13} for quantum expression), abbreviating $\sz_1(t)=\ex{\cLz t}\sz_1$,
\begin{equation}
  D=\lim_{L \to \infty} 4\Gamma L \int_0^\infty \tr{j_{k,k+1} \sz_1(t)}{\rm d}t.
\label{eq:Dj}
  \end{equation}
This expression can be transformed into an alternative form by using the continuity equation for magnetization (see latter derivations), obtaining~\cite{Kamiya13} $D=\lim_{L \to \infty} L \int_0^\infty \tr{j_{k,k+1} \ex{\cLz t}j_{p,p+1}}{\rm d}t$, holding for any $p$ and $k$. By trivially defining the extensive current $J:=L j_{k,k+1}$ the above expression can also be recast into $D=\lim_{L \to \infty} \frac{1}{L} \int_0^\infty \tr{J J(t)}{\rm d}t$, with $J(t):=\ex{\cLz t}J$. Although looking deceptively similar to the standard (equilibrium) Green-Kubo formula~\cite{Pottier} the content is completely different (unitary vs. dissipative evolution).

We now rewrite Eq.(\ref{eq:Dj}) to a form that is better suited for comparison with a unitary setting. Let us denote expectation values in a dissipatively propagated operator $\ex{\cLz t}\sz_1$ as $\z_k^{(0)}(t):=\tr{\sz_k \ex{\cLz t}\sz_1}$ and $j^{(0)}_k(t):=\tr{j_{k,k+1} \ex{\cLz t}\sz_1}$. Taking time derivative and evaluating $\cLz(\sz_1)$, one gets
\begin{equation}
\dot{\z}^{(0)}_1=-4\Gamma \z^{(0)}_1-j^{(0)}_1,\qquad \dot{\z}^{(0)}_L=-4\Gamma \z^{(0)}_L+j^{(0)}_{L-1},
\label{eq:beq}
\end{equation}
while in the bulk one has $\dot{\z}_k^{(0)}=j^{(0)}_{k-1}-j^{(0)}_k$. These are nothing but the continuity equations. The initial condition is $\z_k^{(0)}(0)=\delta_{k,1}$. Integrating (\ref{eq:beq}) over time from 0 to $\infty$, noting that $\z_k^{(0)}(\infty)=0$, one sees that the integral of $j^{(0)}_{L-1}(t)$ needed for $D$ is in turn equal to the integral of $\z_L^{(0)}(t)$, $\int_0^\infty j_k^{(0)}(t){\rm d}t=4\Gamma\int_0^\infty \z_L^{(0)}(t){\rm d}t=1-4\Gamma\int_0^\infty \z_1^{(0)}(t){\rm d}t$. The diffusion constant can therefore be written as
\begin{equation}
D=\lim_{L \to \infty} 16\Gamma^2 L \int_0^\infty\!\!\!\! \tr{\sz_L \sz_1(t)}{\rm d}t,\quad \sz_1(t)=\ex{\cLz t}\sz_1.
\label{eq:Dz}
\end{equation}
In the absence of reflection symmetry $P$ one has to replace $2\tr{\sz_L \sz_1(t)} \to \tr{\sz_L \sz_1(t)}+\tr{\sz_1 \sz_L(t)}$. This equation is our first main result.

It has several nice features. As opposed to the equilibrium Green-Kubo formula, where two limits are necessary, and where in practice for finite (or anomalous) systems an infinite time integral is problematic~\cite{Dhar11,Berciu}, here the time integral always converges regardless of the system size or the transport type (even anomalous) because $\cLz$ is contractive (all nonzero eigenvalues have negative real parts) and $\cLz(\sz_1)\neq 0$. Dissipative dynamics therefore automatically introduces a natural cut-off time given by the inverse of the Lindbladian gap. The only relevant limit to be taken is $L \to \infty$ with the transport type reflected solely in the $L$ dependence of the integral. The NESS current $j=\tr{j_{k,k+1}\rho}$ is an expectation in a complicated NESS $\rho$, while the linear response Eq.(\ref{eq:Dz}) on the other hand gives a more natural interpretation of the same quantity: $D$ is expressed as a transfer probability across the chain, with the evolution $\cLz$ that is unitary except at the boundaries. It suggests that the transport type will be governed by the bulk unitary evolution. Therefore it naturally lends itself to our second goal -- showing the equality of Eq.~(\ref{eq:Dz}) and standard Green-Kubo.

\begin{figure}[ht!]
\centerline{\includegraphics[width=3.2in]{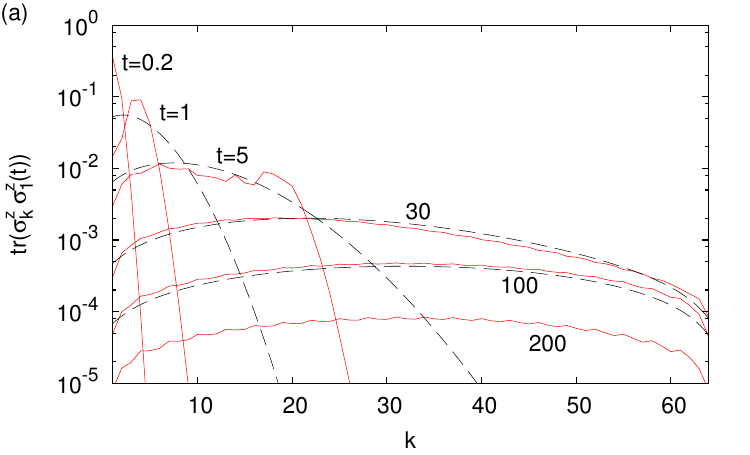}}
\centerline{\hskip3mm\includegraphics[width=3.4in]{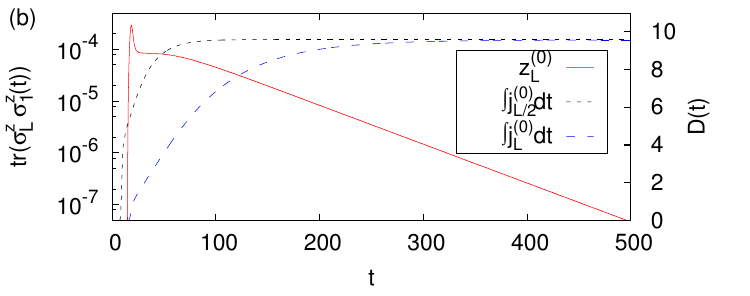}}
\caption{Illustrating NESS Kubo formula (\ref{eq:Dz}) for chaotic XXZ Heisenberg model with $\Delta=0.5$, $h=1$, and $L=64$. (a) magnetization profiles $\tr{\sz_k \ex{\cLz t}\sz_1}$ at selected times (full red curves). Due to unitary bulk evolution magnetization spreads with time from the 1st site and is at the same time leaking out at the boundaries (\ref{eq:beq}). Dashed lines is PDE theory $z(x,t)$ using $\Deq=9.6$ (see the text). (b) Magnetization at the last site (red curve, left axis; its integral gives $D$; at long times it decays with a rate given by the gap of $\cLz$, which scales as $\sim 1/L^{3/2}$), as well as the integral of the current at the middle and the last site (dotted and dashed curves, right axis) again converging at large times to the same $D$, Eq.(\ref{eq:Dj}).}
\label{fig:Fig1}
\end{figure}
Before that let us numerically illustrate Eq.(\ref{eq:Dz}). Taking the Heisenberg XXZ chain in a staggered field, $H=\sum_j \sx_j \sx_{j+1}+\sy_j \sy_{j+1}+\Delta \sz_j\sz_{j+1}+\frac{1}{2}(h_j \sz_j+h_{j+1}\sz_{j+1})$, with $h_{3k}=-h, h_{3k+1}=-h/2, h_{3j+2}=0$, one has a quantum chaotic model (random matrix level spacing statistics~\cite{PRE10}) for which diffusion is expected. We numerically (see Appendix) evaluate different expectations in $\ex{\cLz t}\sz_1$, shown in Fig.~\ref{fig:Fig1}. The initial magnetization spreads from site $1$, with corresponding integrals resulting in $D$.

\section{Equality of diffusion}
Looking at Eq.(\ref{eq:Dz}) it is not clear that it gives the same $D$ as the equilibrium Green-Kubo formula. For example, naively $D$ looks proportional to $\Gamma^2$ (a dependence on $\Gamma$ has indeed been observed in small systems~\cite{Berciu}). Our aim is to show rigorously and in general that, provided the unitary dynamics (i.e., $H$) is diffusive, the transport coefficient obtained by (\ref{eq:Dz}) is the same as the unitary $\Deq$.

To show this we use exact conservation equations at the boundary (\ref{eq:beq}) while we replace a complicated evolution equation of the current $j_k^{(0)}$ by a simpler one, assuming that the Fick's law holds, $j_k^{(0)}=-\Deq (\z_{k+1}^{(0)}-\z_{k}^{(0)})$. This is to say that the dissipative part of $\cLz$ is treated exactly while the unitary evolution in the bulk is assumed to be perfectly diffusive. Here we specifically stress that $\Deq$ is the unitary diffusion coefficient of bulk dynamics (e.g., obtained from the Green-Kubo formula) which could be different than the NESS one $D$ (\ref{eq:Dz}) for any of the mentioned reasons (``unrealistic'' driving, boundary driving modifying dynamics, etc.). We show that this is not the case. Fick's law in the bulk together with (\ref{eq:beq}) constitutes a closed set of $L$ coupled differential equations for $\z_k^{(0)}(t)$, which are nothing but a discrete diffusion equation $\dot{\z}_k^{(0)}=\Deq (\z_{k+1}^{(0)}+\z_{k-1}^{(0)}-2\z_k^{(0)})$ plus a dissipative boundary condition (\ref{eq:beq}). We are especially interested in the large-$L$ behavior where we write a partial differential equation (PDE) for $z(x,t)$, $\dot{z}(x,t)=\Deq z''(x,t)$, with boundary conditions,
\begin{eqnarray}
\dot{z}(0,t)&=&-4\Gamma z(0,t)-\Deq z'(0,t)\nonumber \\
\dot{z}(L,t)&=&-4\Gamma z(L,t)+\Deq z'(L,t),
\label{eq:PDE}
\end{eqnarray}
and the initial condition $z(x,0)=\delta(x-0^+)$. Absorbing boundary conditions (\ref{eq:PDE}) result in a slightly non-standard problem that can nevertheless be solved by a separation of variables. Writing the solution in terms of eigenfunctions $X_n(x)$ as $z(x,t)=\sum_n c_n X_n(x) \ex{-\Deq k_n^2 t}$, we get (see Appendix~\ref{app2})
\begin{equation}
X_n(x)=\cos{(k_n x)}+\frac{4\Gamma-\Deq k_n^2}{\Deq k_n}\sin{(k_n x)},
\end{equation}
with a transcendental eigenvalue equation for $k_n$,
\begin{equation}
\tan{(k_n L)}=-2\Deq k_n \frac{(4\Gamma-\Deq k_n^2)}{(4\Gamma-\Deq k_n^2)^2-\Deq^2 k_n^2}.
\label{eq:kn}
\end{equation}
$X_n$ are orthogonal with respect to a modified inner product $\inn{X_n}{X_m} :=\int_0^L X_n(x) X_m(x){\rm d}x+X_n(0)X_m(0)+X_n(L)X_m(L)$. The initial condition gives $c_n=\frac{1}{\inn{X_n}{X_n}}$. We can now express finite-$L$ NESS $D$ (\ref{eq:Dz}) as
\begin{equation}
D=16\Gamma^2 L \int_0^\infty\!\!\!\!\! z(L,t){\rm d}t=\frac{16\Gamma^2 L}{\Deq}\sum_{n=1}^\infty\frac{-(-1)^n}{k_n^2 \inn{X_n}{X_n}}.
\label{eq:sum}
\end{equation}
In the TDL one can replace the sum with an integral (we checked, see Appendix~\ref{app2}, that this describes the exact sum (\ref{eq:sum}) well even for not so large $L \sim 16$), resulting in
\begin{equation}
D=\frac{\Deq}{1+\frac{\Deq}{2\Gamma L}}\approx\Deq(1-\frac{\Deq}{2\Gamma L}).
\label{eq:DL}
\end{equation}
This is our second main result.

The linear response NESS transport coefficient $D$ (\ref{eq:Dz}), defined via NESS current scaling (\ref{eq:NESSF}), is in the leading order in $L$ {\em exactly equal} to the bulk unitary transport coefficient $\Deq$. Furthermore, finite size corrections should scale as $\sim 1/L$. For weak driving $\mu$ and fixed coupling $\Gamma$ one always has $D=\Deq$ in the TDL. The only assumption going into deriving this result is that in bulk, where one has only unitary evolution, Fick's law holds. If Fick's law holds only on some hydrodynamic length-scale of $l_*$ lattice spacings we expect that the above expression changes to
\begin{equation}
D \asymp \Deq\left(1-\frac{\alpha(\Gamma)}{ (L/l_*)}\right),
\label{eq:Dg}
\end{equation}
with possibly complicated $\alpha(\Gamma)$ that is not necessarily $1/\Gamma$. If the Fick's law $\Deq$ has subleading corrections in $L$ (either due to a boundary, or due to bulk dynamics) this can modify the convergence of $D$, however, one will still have $D=\Deq$ in the TDL. The correct order of limits does matter: if one takes a fixed $L$ and $\Gamma \to 0$ the diffusion constant goes to zero (see Appendix~\ref{app2}); if one takes first $\Gamma \to 0$ and only then weak driving  $\mu \to 0$ and $L \to \infty$ the diffusion constant diverges~\cite{Prosen11}.

Let us test the result (\ref{eq:DL}) on three microscopic models. XX chain with bulk dephasing is a non-quadratic exactly solvable diffusive model in a single-particle~\cite{Esposito05}  as well as in a many-particle~\cite{JSTAT10} situation, with an exact expression~\cite{JSTAT10} for the NESS $D:=\frac{j(L-1)}{2\mu}$ being $D=\Deq/(1+\frac{\Deq(\Gamma+1/\Gamma)}{2(L-1)})$, where we defined $\Deq:=\lim_{L \to \infty}j L/2\mu=2/\gamma$~\cite{JSTAT10,Hartnoll}. For small $\Gamma$ this is exactly the same as the above general relation (\ref{eq:DL}). 
\begin{figure}[t!]
\centerline{\includegraphics[width=3.2in]{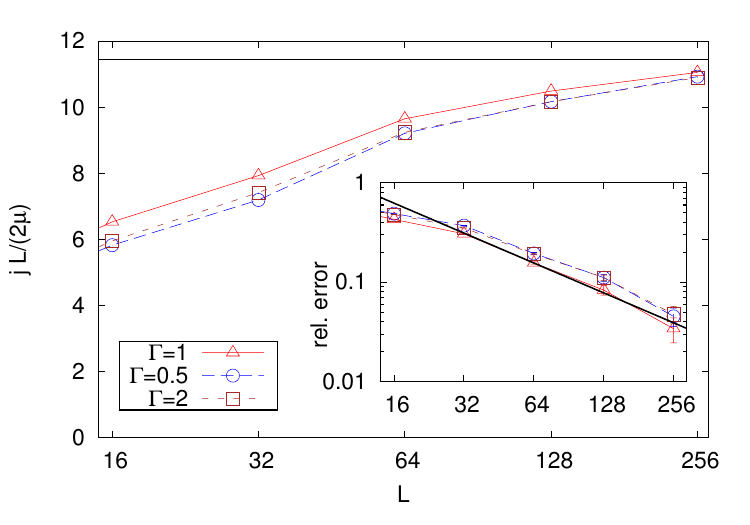}}
\caption{Convergence of NESS $D$ with system size $L$ in the chaotic staggered XXZ Heisenberg model ($h=1, \Delta=0.5, \mu=0.02$). Horizontal line is the asymptotic value $D \approx 11.45$. The inset shows convergence of $1-D(L)/D(\infty)$ (full line is $10/L$).}
\label{fig:chao}
\end{figure}
Next, we take the chaotic staggered XXZ model. In Fig.~\ref{fig:chao} we see that the finite-size correction indeed scales as $1/L$, however, the dependence on $\Gamma$ is not as in Eq.(\ref{eq:DL}) but rather more general (\ref{eq:Dg}). We can see in Fig.~\ref{fig:Fig1} (dashed curves) that the solution $z(x,t)$ of the PDE (\ref{eq:PDE}) describes full quantum evolution rather well at longer times when diffusion emerges. Lastly, we take the integrable XXZ chain with $h=0$ and $\Delta=1.5$ at half-filling, where previous results indicate high-temperature diffusion, see e.g. Refs.~\cite{sachdev,Long04,JSTAT09,sirker,Karrasch:14,Stein:09,Stein:12,Ours17}. Our data show (Appendix~\ref{app3}) that convergence is in this case not $\sim 1/L$ as predicted for diffusive systems (\ref{eq:Dg}), but rather slower $\sim 1/L^\alpha$ with the power around $\alpha \approx 0.5$ (see also data in the Supplement of Ref.~\onlinecite{PNAS18} for similar slow convergence in a different model). Significance of that is at present not clear (Appendix~\ref{app3}).

\section{Conclusion}
Studying nonequilibrium steady state physics of 1D quantum systems, focusing on high-temperature particle (magnetization) transport, we derive a weak driving nonequilibrium Kubo-like expression for the diffusion constant. It has some advantages over the equilibrium Green-Kubo formula and lends itself to comparison with unitary transport calculation. Without any further assumptions we show that provided the unitary dynamics is diffusive (Fick's law is valid) the nonequilibrium formula gives exactly the same diffusion constant as the equilibrium Green-Kubo formula. We also predict a universal $\sim 1/L$ convergence with system size. While the result is derived for a specific quantum boundary driving, it could be generalized to any boundary driven NESS setting, including e.g. classical stochastic models~\cite{Derrida}. The nonequilibrium Kubo formula should be of wide use in transport studies of diffusive as well as anomalous many-body systems.

\section*{Acknowledgements}
I would like to thank T.~Prosen and L.~Zadnik for discussion and acknowledge Grants No.~J1-7279 and No.~P1-0044 from Slovenian Research Agency.

\appendix

\section{Lindbladian perturbation theory}
\label{app1}

Let us write the Lindbladian as a sum of two linear operators (in the examples $\cLz$ is also Lindbladian while $\cLe$ is only linear but not Lindbladian),
\begin{equation}
\cL=\cLz+\mu \cLe,
\end{equation}
where $\mu$ is some small parameter. The (unique) steady state of $\cLz$ is denoted by $\rho_0$, $\cLz \rho_0=0$. For small $\mu$ we look for a perturbative solution
\begin{equation}
\rho=\rho_0 + \mu \rho_1+\cdots,
\end{equation}
getting a standard perturbation theory expression for the steady-state linear correction $\rho_1$,
\begin{equation}
\cLz \rho_1=-\cLe \rho_0=:-R,
\end{equation}
where we defined $R:=\cLe \rho_0$. Formally, one can write 
\begin{equation}
\rho_1=-\cLz^{-1}(R).
\label{eq:SinvL0}
\end{equation}
This expression is well defined (has a unique solution) provided $R$ is orthogonal to the kernel of $\cLz$, in other words, if $\cLe \rho_0$ is orthogonal to $\rho_0$ (this holds true for cases of interest discussed latter).

Alternatively, one can write the linear-response equation for a time-dependent perturbation $\rho_1(t)$,
\begin{equation}
\dot{\rho}_1(t)=\cLz \rho_1+\cLe \rho_0,
\end{equation}
which is a linear inhomogeneous equation for $\rho_1(t)$. The formal solution satisfying $\rho_1(0)=0$ is $\rho_1(t)=\int_0^t \ex{\cLz\cdot(t-\tau)}R{\rm d}\tau$, where $R:=\cLe \rho_0$. The steady-state correction can therefore also be written as~\cite{Michel04}
\begin{equation}
\rho_1=\rho_1(t \to \infty)=\int_0^\infty \ex{\cLz \tau}R{\rm d}\tau=\int_0^\infty R(\tau){\rm d}\tau,
\end{equation}
which is a formal way of writing the (pseudo)inverse in Eq.(\ref{eq:SinvL0}). Note that $R(t)=\ex{\cLz t}R$ goes to zero (in any norm) at long times because of contractivity of $\cLz$ and the fact that $R$ is orthogonal to the kernel of $\cLz$. Even in a finite system the integral therefore converges regardless of the dynamics.

\section{Solving the PDE}
\label{app2}

\begin{figure}[t!]
\centerline{\includegraphics[width=3in]{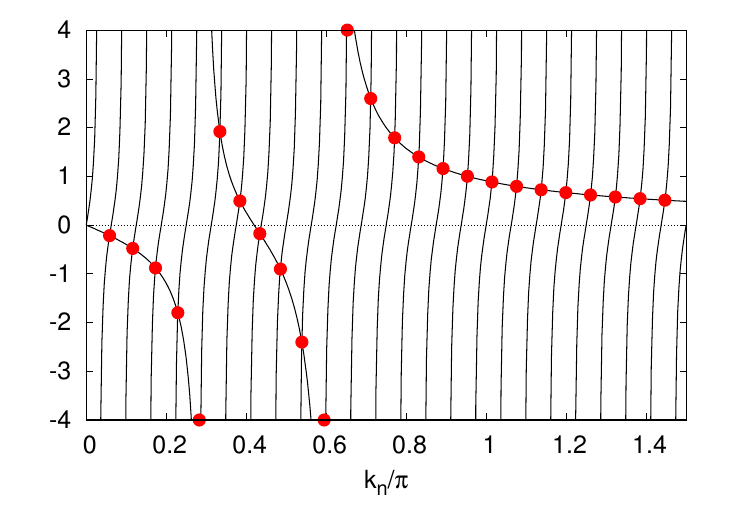}}
\caption{Solutions of Eq.(\ref{eq:Skn}) (red points) for $\Deq=2.3$, $\Gamma=1$ and $L=16$. Two sets of curves are left and right-hand sides of Eq.(\ref{eq:Skn}).}
\label{fig:kn}
\end{figure}
We solve for time evolution by $\cLz$ by using exact dissipative boundary conditions while for a constitutive relation that connects local current to other local observables (like magnetization), and which is in principle complicated and depends on the specifics of each $H$, we take the Fick's law,
\begin{equation}
j_k^{(0)}=-\Deq (\z_{k+1}^{(0)}-\z_{k}^{(0)}).
\end{equation}
This makes for a close set of equations for magnetizations $\z_k^{(0)}$. In the continuum limit we can replace a set of $L$ coupled differential equations by a PDE. Namely, we want to solve (a dot denotes time derivatives, primes denote spatial derivatives)
\begin{equation}
\dot{z}(x,t)=\Deq z''(x,t),
\end{equation}
with boundary conditions,
\begin{eqnarray}
\dot{z}(0,t)&=&-4\Gamma z(0,t)-\Deq z'(0,t)\\
\dot{z}(L,t)&=&-4\Gamma z(L,t)+\Deq z'(L,t), \nonumber
\end{eqnarray}
and the initial condition $z(x,0)=\delta(x-0^+)$. We write the solution as 
\begin{equation}
z(x,t)=\sum_n c_n X_n(x) \ex{-\Deq k_n^2 t},
\end{equation} 
in terms of eigenfunctions $X_n(x)$ satisfying the eigenequation $X_n''+k_n^2 X_n=0$. Eigenfunctions are $X_n(x)=A\cos{(k_n x)}+B \sin{(k_n x)}$ and have to satisfy boundary conditions $(4\Gamma-\Deq k_n^2)X_n(0)-\Deq X_n'(0)=0$ and $(4\Gamma-\Deq k_n^2)X_n(L)+\Deq X_n'(L)=0$. Choosing $A=1$ and $B=(4\Gamma-\Deq k_n^2)/(\Deq k_n)$ satisfies the first boundary condition, so that the unnormalized eigenfunctions are
\begin{equation}
X_n(x)=\cos{(k_n x)}+\frac{4\Gamma-\Deq k_n^2}{\Deq k_n}\sin{(k_n x)},
\label{eq:SXn}
\end{equation}
while the second one leads to a transcendental equation for eigenvalues $k_n$,
\begin{equation}
\tan{(k_n L)}=-2\Deq k_n \frac{(4\Gamma-\Deq k_n^2)}{(4\Gamma-\Deq k_n^2)^2-\Deq^2 k_n^2}.
\label{eq:Skn}
\end{equation}
See Fig.~\ref{fig:kn} for an illustration.

\begin{figure}[t!]
\centerline{\includegraphics[width=2.9in]{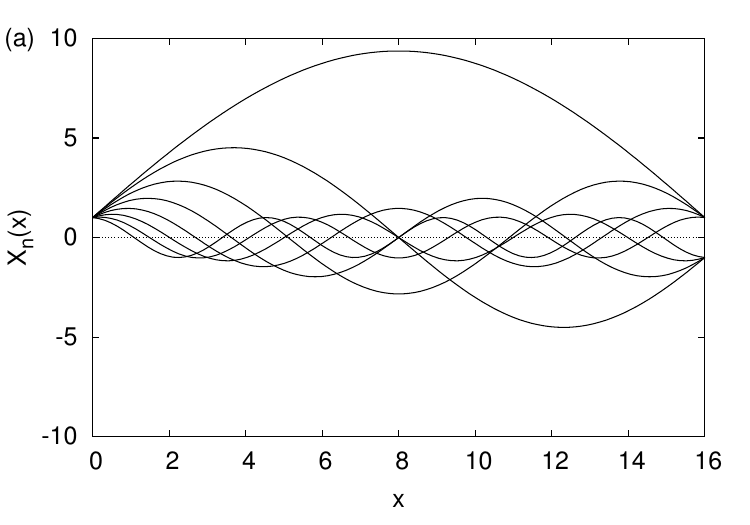}}
\centerline{\includegraphics[width=2.9in]{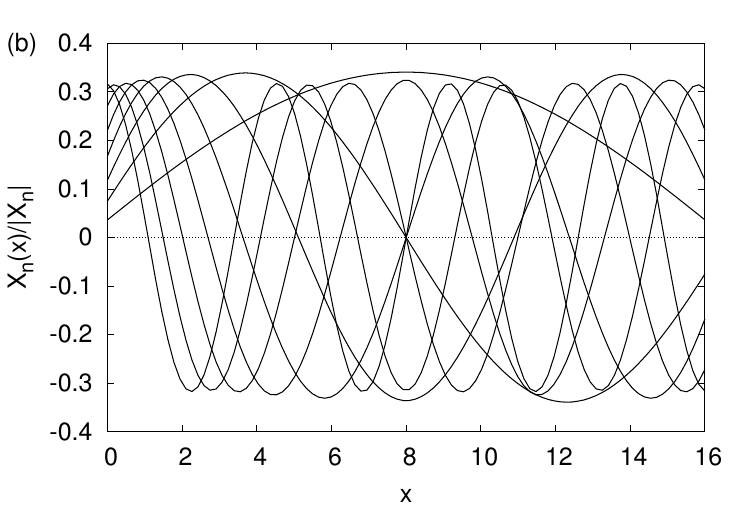}}
\caption{First eight eigenfunctions $X_n(x)$ (\ref{eq:SXn}). (a) shows unnormalized and (b) normalized eigenfunctions, both for $\Gamma=1$, $\Deq=2.3$ and $L=16$.}
\label{fig:Xn}
\end{figure}

Because the boundary conditions depend on the eigenvalue $k_n$ one gets a modified inner product (it is not one of the usual, simpler, Sturm-Liouville homogeneous boundary conditions with fixed coefficients). Using standard procedure, multiplying the eigenequation for $X_n$ by $X_m$, integrating over $x$ and making one per-partes integration, one ends up with $(k_n^2-k_m^2)\inn{X_n}{X_m}=0$, leading to the orthogonality of $X_n$ with respect to the inner product defined as,
\begin{eqnarray}
\inn{X_n}{X_m} :=&&\int_0^L X_n(x) X_m(x){\rm d}x+ \\
&&+X_n(0)X_m(0)+X_n(L)X_m(L). \nonumber
\end{eqnarray}
The initial condition in turn fixes the expansion coefficients $c_n$ to simple $c_n=1/\inn{X_n}{X_n}$ because one always has $X_n(0)=1$. At the other end one has $X_n(L)=(-1)^{n+1}$. See Fig.~\ref{fig:Xn} for an example of few eigenfunctions.

\begin{figure}[t!]
\centerline{\includegraphics[width=3in]{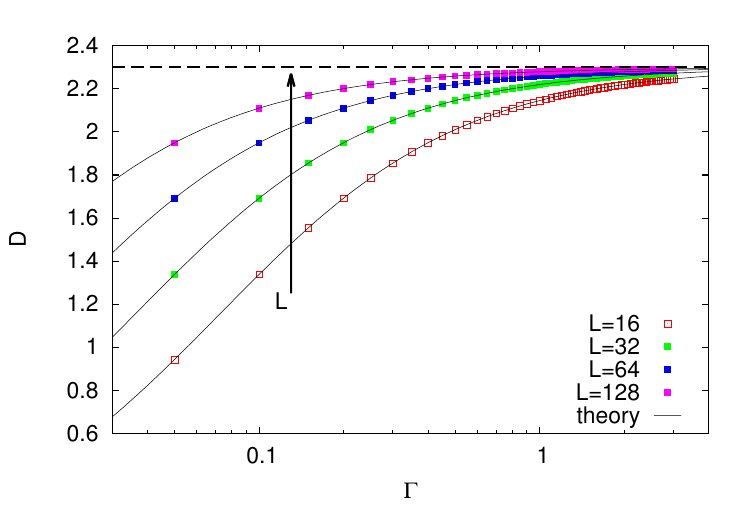}}
\caption{Comparison of the NESS diffusion coefficient $D$ obtained from the exact sum (\ref{eq:Ssum}) over eigenvalues $k_n$ satisfying (\ref{eq:Skn}) (symbols; we use the lowest $\sim 8L$ eigenvalues) and continuum theory (full curves, Eq.(\ref{eq:SDL}). Already for small $L$ Eq.(\ref{eq:SDL}) obtained by replacing the sum with an integral describes the dependence perfectly. At fixed coupling strength $\Gamma$ and increasing $L$ the NESS diffusion constant $D$ converges to $\Deq=2.3$.}
\label{fig:cmp}
\end{figure}
We can now express the NESS finite-$L$ diffusion constant (\ref{eq:Dz}) as
\begin{equation}
D=16\Gamma^2 L \int_0^\infty\!\!\! z(L,t){\rm d}t=\frac{16\Gamma^2 L}{\Deq}\sum_{n=1}^\infty\frac{-(-1)^n}{k_n^2 \inn{X_n}{X_n}},
\label{eq:Ssum}
\end{equation}
where $k_n$ are solutions of Eq.(\ref{eq:kn}). The norm of $X_n$ can be evaluated, and is after simplification (taking into account (\ref{eq:Skn})),
\begin{equation}
\inn{X_n}{X_n}=\frac{L}{2}\left(1+\frac{(4\Gamma-k_n^2 \Deq)^2}{k_n^2 \Deq^2}\right)+1+\frac{4\Gamma}{\Deq k_n^2}.
\label{eq:Snorm}
\end{equation}
Denoting $f(k_n):=\frac{1}{k_n^2 \inn{X_n}{X_n}}$, in the limit of large $L$, when $k_n \approx n\frac{\pi}{L}$, we are dealing with a sum (\ref{eq:Ssum}) of terms like $f(n \pi/L)-f((n+1)\pi/L)\approx -f'(k)\pi/L$. Replacing the sum with an integral one gets
\begin{equation}
D=\frac{16\Gamma^2 L}{\Deq} \int_0^\infty \frac{-f'(k)}{2}{\rm d}k.
\end{equation}
Despite a complicated $f'(k)$ the integral can nevertheless be evaluated in a closed form, resulting in
\begin{equation}
D=\frac{\Deq}{1+\frac{\Deq}{2\Gamma L}}.
\label{eq:SDL}
\end{equation}
In Fig.~\ref{fig:cmp} we compare the continuum formula (\ref{eq:SDL}) and the exact sum (\ref{eq:Ssum}), seeing that the replacement of a sum with an integral gives good results already for small $L=16$.

It is instructive to understand where does the $\sim 1/L$ correction in $D$ come from. It is due to the last term in the norm (\ref{eq:Snorm}), namely, due to $\frac{4\Gamma}{\Deq k_n^2}$. In the norm (\ref{eq:Snorm}) the first term, proportional to $L$, is simply due to the length of the interval while the last, $L$-independent $4\Gamma/\Deq k_n^2$, is due to the fact that one does not have an integer number of oscillations in $x \in [0,L]$ (see Fig.~\ref{fig:Xn}). For instance, integrating $\cos^2{(k_n x)}=(1+\cos{(2k_n x)})/2$ one gets ``boundary'' terms like $\sin{(2k_n L)}$. In other words, the last term responsible for $\sim 1/L$ correction is due to the boundary condition that causes a ``phase shift'' such that the boundary condition $X_n({0,L})=\pm 1$ is satisfied. Writing this term as $\frac{8a}{k_n^2}$ one would get $\frac{\Deq}{D}=1+\frac{a \Deq^2}{\Gamma^2 L}$. The stronger the effect of the boundary, i.e., the larger $a$, the larger is finite-size correction.

\section{Microscopic XXZ model}
\label{app3}

\begin{figure}[htb]
\centerline{\includegraphics[width=3in]{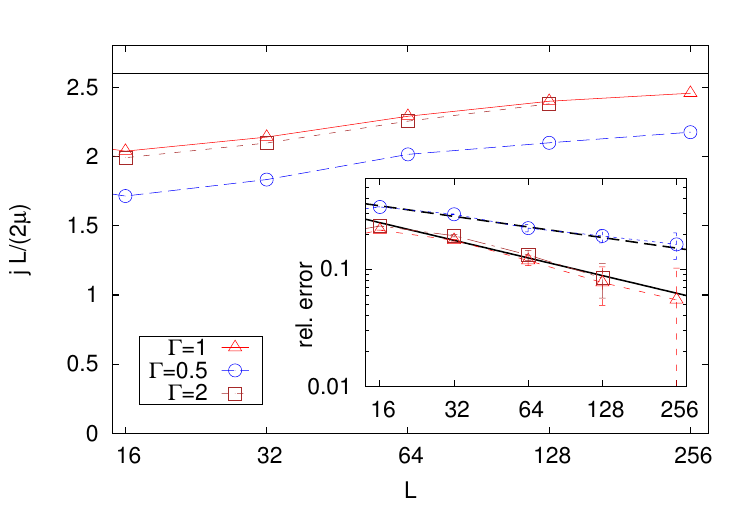}}
\caption{Convergence of the NESS diffusion constant with $L$ for the integrable XXZ Heisenberg chain with $\Delta=1.5$ ($h=0$). Full line is the asymptotic value $D(L \to \infty)\approx 2.6$. The inset shows relative error at finite $L$, i.e. $1-D(L)/D(\infty)$, that here decays slower than predicted for diffusive theory (\ref{eq:Dg}). Namely, two black lines are $1/L^{0.5}$ (full) and $0.8/L^{0.3}$ (dashed).}
\label{fig:XXZ}
\end{figure}

Using time-dependent density-matrix renormalization group (tDMRG) method and the mentioned Lindblad magnetization driving we study spin transport in a class of XXZ spin chains,
\begin{equation}
H=\sum_{j=1}^{L-1} \sx_j \sx_{j+1}+\sy_j \sy_{j+1}+\Delta \sz_j\sz_{j+1}+\frac{1}{2}(h_j \sz_j+h_{j+1}\sz_{j+1}),
\end{equation}
with $h_{3k}=-h, h_{3k+1}=-h/2, h_{3j+2}=0$. For $h=1$ we have quantum chaotic model~\cite{PRE10}, while for $h=0$ the model is integrable. Spin (magnetization) current operator is $j_{k,k+1}=2(\sx_k \sy_{k+1}-\sy_k \sx_{k+1})$. For small driving $\mu$, we typically use $\mu=0.01$, the NESS is close to the identity operator and one therefore studies infinite-temperature transport at half-filling (zero magnetization). Details of numerical implementation can be found in e.g.~\cite{JSTAT09,PRL16} and references cited therein.

In the main text we presented data for a chaotic system, here we study the integrable case obtained for $h=0$ and $\Delta=1.5$, where diffusion was observed. Indeed, we see (Fig.~\ref{fig:XXZ}) that with system size $D$ converges to a constant independent of $\Gamma$. However, the convergence is slower. Finite-size correction does not scale as $\sim 1/L$, predicted by our theory for diffusive bulk evolution, but rather as $\sim 1/L^\alpha$ with $\alpha \approx 0.5$ for $\Gamma=1$ (precise value is hard to determine due to limited $L$). We do not at present understand the origin of such slow convergence. Remember that $\sim 1/L$ correction in the case of diffusion was due to boundary effects, which in a diffusive system are expected to have a finite extent around the edge. Stronger finite-size effects, like $1/L^{0.5}$, could either suggest that the effect of a boundary extends further into the system (it should affect $\sim L^{0.5}$ sites), or that the Fick's law has $\sim 1/L^{0.5}$ corrections in the bulk. It is not clear if it signals some non-diffusive physics; we note that in higher NESS current fluctuations non-diffusive scaling has indeed been observed~\cite{PRB14}. What is puzzling is that similar slow convergence has also been observed in a weakly perturbed XXZ model~\cite{PNAS18} (which is not integrable anymore), so it could be an effect having an origin in some particular property of the XXZ model. An alternative explanation could also be that in the XXZ model finite size effects are simply larger, and at $L=256$ we might not yet be in the asymptotic regime of $\sim 1/L$ scaling (magnetization profiles though are nicely linear for studied sizes).


\begin{thebibliography}{99}
\bibitem{foot0} An argument why such driving might be ``unphysical'' is that it is hard to derive it from a realistic microscopic $H$. Namely, starting from a Hamiltonian of a bath and a system, a standard derivation of the Lindblad equation~\cite{breuer} requires among other things weak coupling, and results in a weakly coupled nonlocal $L_j$ (see though e.g. Ref.~\cite{platini} for a ``repeated interaction'' picture of local driving). Driving used is on the other hand local and strong ($\Gamma \sim 1$), and we know~\cite{Prosen11} that having weak local coupling is not the way to go as it probes non-bulk physics. Our results show that such objections are indeed irrelevant for bulk physics in the TDL.
\bibitem{foot1} Only in the TDL can one unambiguously distingush different transport types differing in e.g. scaling $x^2 \sim t^\alpha$.
\bibitem{foot2} Thermalization is also required for the validity of equilibrium Green-Kubo formulas. Our focus is on interacting many-body systems in the TDL where thermalization is expected to be generic.
\end{thebibliography}
\end{document}